\documentclass{emulateapj}

\usepackage{natbib} 
\begin{document}

\title{The DEEP2 Galaxy Redshift Survey: Probing the Evolution of Dark Matter
Halos around Isolated Galaxies from $z\sim1$ to $z\sim0$}

\author{ Charlie Conroy\altaffilmark{1},  Jeffrey
A. Newman\altaffilmark{2},  Marc Davis\altaffilmark{1,3},  Alison
L. Coil\altaffilmark{1},  Renbin Yan\altaffilmark{1} Michael
C. Cooper\altaffilmark{1} Brian F. Gerke\altaffilmark{3},
S.~M. Faber\altaffilmark{4}, David C. Koo\altaffilmark{4} }

\altaffiltext{1}{Department of Astronomy, University of California,
Berkeley, CA 94720 -- 3411} 
\altaffiltext{2}{Hubble Fellow, Lawrence Berkeley National Laboratory, 
1 Cyclotron Road, Berkeley, CA 94720}
\altaffiltext{3}{Department of Physics, University of California,
Berkeley, CA 94720 -- 3411}
\altaffiltext{4}{University of California Observatories/Lick
Observatory, Department of Astronomy and Astrophysics, University of
California, Santa Cruz, CA 95064}

\begin{abstract}
Using the first 25\% of DEEP2 Redshift Survey data, we probe the
line-of-sight velocity dispersion profile for isolated galaxies with
absolute B-band magnitude $-22<M_{B}-5\log(h)<-21$ at $z$=0.7-1.0,
using satellite galaxies as luminous tracers of the underlying
velocity distribution.  Measuring the velocity dispersion beyond a
galactocentric radius of $\sim200 h^{-1}$kpc (physical) permits us to
determine the total mass, including dark matter, around these bright
galaxies.  Tests with mock catalogs based on N-body simulations
indicate that this mass measurement method is robust to selection
effects.  We find a line-of-sight velocity dispersion ($\sigma_{los}$)
of $162^{+44}_{-30}$ km s$^{-1}$ at $\sim110 h^{-1}$ kpc,
$136^{+26}_{-20}$ km s$^{-1}$ at $\sim230 h^{-1}$ kpc, and
$150^{+55}_{-38}$ km s$^{-1}$ at $\sim320 h^{-1}$ kpc.  Assuming an
NFW model for the dark matter density profile, this corresponds to a
mass within r$_{200}$ of $M_{200}=5.5^{+2.5}_{-2.0}\times10^{12}
h^{-1}$M$_{\Sun}$ for our sample of satellite hosts with mean
luminosity $\sim2.5L^\ast$.  Roughly $\sim60\%$ of these host galaxies
have early-type spectra and are red in restframe $(U-B)$ color,
consistent with the overall DEEP2 sample in the same luminosity and
redshift range.  The halo mass determined for DEEP2 host galaxies is
consistent with that measured in the Sloan Digital Sky Survey for host
galaxies within a similar luminosity range relative to $M^\ast_B$.
This comparison is insensitive to the assumed halo mass profile, and
implies an increase in the dynamical mass-to-light ratio
($M_{200}/L_B$) of isolated galaxies which host satellites by a factor
of $\sim 2.5$ from $z\sim1$ to $z\sim0$.  Our results can be used to
constrain the halo occupation distribution and the conditional
luminosity function used to populate dark matter halos with galaxies.
In particular, our results are consistent with scenarios in which
galaxies populate dark matter halos similarly from $z\sim 0$ to $z\sim
1$, except for $\sim 1$ magnitude of evolution in the luminosity of
all galaxies.  With the full DEEP2 sample, it will  be possible to
extend this analysis to multiple luminosity or color bins.

\end{abstract}


\section{Introduction}\label{s:intro}

It has been firmly established that galaxies and clusters form within
halos whose mass is dominated by unseen dark matter.  Yet until very
recently, the outer regions of halos have been very poorly understood
due to a lack of visible tracers of the mass distribution.
Galaxy-galaxy lensing is able to probe the halo masses of local
galaxies, though with some degree of uncertainty, as this method
actually  probes all of the mass along the line-of-sight
\citep{Guzik02}, and has  only recently been applied to isolated
galaxies \citep{Hoekstra05}.  Recent work \citep{Wilson01, Hoekstra04,
Kleinheinrich05} suggests that the virial  mass of $\sim L^\ast$
galaxies has remained constant from $z\sim0.8$ to $z\sim0.15$.  Beyond
$z\sim0.5$ the lensing probability rapidly diminishes, making it very
difficult to derive masses of isolated  galaxies (with masses
$\sim10^{12} M_{\sun}$) with this method at higher redshift \citep[see
e.g.,][]{Peacock99}.

The dynamics of satellite galaxies orbiting larger ``host'' galaxies
provide another way to probe the mass distribution at large radii.
Early work by \citet{LT87,EGH87,Zar89} utilized samples of tens of
satellites as early confirmation that galaxies are embedded in large
dark matter halos.  More conclusive evidence was compiled by
\citet{zar93,zar97} who used the kinematics of a sample of 115
satellite galaxies to probe the outer regions of 69 isolated galaxies.
By employing satellites as test particles, they built up a velocity
profile for a single representative isolated galaxy by stacking
measurements of satellites from many different host galaxies.

More recently, \citet[hereafter P03]{P03} use $\sim$250,000 SDSS
redshifts to probe the halo masses of isolated galaxies; they detect
$>$1000 satellites around $\sim$700 host galaxies.  With this large
data set they are able to discriminate between various halo mass
distributions and find evidence for an NFW-like falloff ($\rho \propto
r^{-3}$) at large radii.  From these accurate line-of-sight velocity
dispersion profile measurements, P03 infer the masses enclosed within
1.5 R$_{virial}$ for two sets of isolated galaxies. Host galaxies with
$-20.5<M_{B}<-19.5$ are found to have an average halo mass of
$M_{virial}\approx1.5\times10^{12} M_{\Sun}$ while hosts with
$-21.5<M_{B}<-20.5$ have $M_{virial}\approx6\times10^{12} M_{\Sun}$
(for $h=0.7$).  Other recent work utilizing satellite dynamics
includes \citet{McKay02}, who check SDSS weak-lensing scaling laws;
\citet{VDB04b,VDB04a}, who use mock galaxy catalogs and the 2dF survey
to constrain the conditional luminosity function and investigate the
levels and effects of contamination in dynamical satellite studies;
and \citet{brainerd05}, who measure velocity dispersion  profiles for
subsamples of isolated 2dF galaxies.

By extending this type of measurement to high redshift, we can study
the evolution of the relationship between galaxies and dark matter
halos.  There have been few ways to do this until the recent advent of
large, high-redshift surveys.  The best example to date is
\citet{Yan03b}, who use the 2dF and DEEP2 two-point correlation
functions \citep[][respectively]{Madg03c, coil04} to constrain the
evolution of the halo occupation distribution, a key ingredient  of
the halo model.  Yan et al. produced a set of $z\sim 1$ mock catalogs
using N-body simulations and a halo model whose parameters are set by
requiring a fit to $\xi(r)$ from 2dF at $z\sim 0$.  They find good
agreement between the correlation statistics at $z\sim 0.8$ from DEEP2
and the prediction from these mock catalogs, which populate galaxies
in dark matter halos in the same way (as a function of $L/L^\ast$ and
halo mass) at $z\sim 1$ and $z\sim 0$.
Hence their results are consistent with a minimal-evolution
hypothesis, in which galaxies with a given luminosity compared to
$L^\ast$ at $z\sim 1$ live in the same sorts of halos as similar
galaxies at $z\sim0$, though the mass function of dark matter halos
and $L^\ast$ evolve.  Here we address this hypothesis with an
independent method.

In this paper we constrain the velocity dispersion profile for a
typical isolated DEEP2 galaxy at $z\sim0.8$ using methods similar to
those of P03.   We then deduce a representative halo mass for these
galaxies and compare our  results to recent local measurements from
SDSS to test for  evolution in a self-consistent way.  We use mock
catalogs to test the significance and  robustness of these results.
The paper proceeds as follows.   In $\S$~\ref{s:data} we describe the
DEEP2 data set and the properties of  satellite galaxies and their
hosts.  $\S$~\ref{s:meth} outlines our method for  reconstructing the
mass of isolated galaxies using satellites and in $\S$~\ref{s:res} we
present our results and compare with recent local measurements.  We
test our methods using mock catalogs in $\S$~\ref{s:mocks} and discuss
some implications of our results in $\S$~\ref{s:disc}.  Throughout the
paper we assume a standard $\Lambda$CDM cosmology with
$\Omega_{m}=0.3$, $\Omega_{\Lambda}=0.7$ and H$_{0}=100h^{-1}$ km
s$^{-1}$ Mpc$^{-1}$.  Absolute magnitudes quoted have been K-corrected
and corrected for reddening  by galactic dust, and are in the AB
system \citep{willmer05}.

\section{Satellite-Host Systems at $z\sim1$}\label{s:data}

In this Section we introduce the data used at $z\sim1$, describe the
algorithm used to  identify bright isolated galaxies and their
satellites, and highlight several  properties of these host-satellite
systems.

We use data from the first $\sim25\%$ of the DEEP2 Galaxy Redshift
Survey, a three-year project using the DEIMOS spectrograph at the 10-m
Keck II telescope  to survey galaxies at $z\sim1$.  DEEP2 will collect
spectra  of $\sim$50,000 galaxies from $0.7<z<1.4$ to a limiting
magnitude of $R_{AB} = 24.1$ with redshift errors of $\sim$20 km
s$^{-1}$.  For survey details,  see \citet{Davis04}.  Photometric data
were taken in the $B, R$ and $I$ bands with the CFH12k camera on the
3.6-m Canada-France-Hawaii telescope.  We use data taken during the
first two seasons  of DEEP2 observations, which have yielded
$\sim$12,000 secure  redshifts over $\sim$0.9 sq. degrees.   Our
observed R-band limiting magnitude corresponds to a different
restframe wavelength with redshift, from $4000\AA$ at $z=0.7$ to
$2800\AA$ at $z=1.4$.  This results in a different selection function
for red and blue galaxies with redshift, such that as we move to
fainter magnitudes, red galaxies  become undetectable before blue
galaxies; this effect increases with  increasing redshift
\citep[see][for details]{willmer05}.  To minimize  this effect we
consider only bright hosts with $z<1$.

We define an isolated galaxy as having no bright neighbors within a
given  search cylinder.  Once isolated galaxies have been identified,
we use another search  cylinder to identify faint satellite
companions.  Specifically, a galaxy is isolated if it has no neighbors
in the DEEP2 spectroscopic sample within a physical distance projected
on the sky $r_{p}<350 h^{-1}$kpc, line-of-sight velocity difference
$|\Delta v|<1000$ km s$^{-1}$ and absolute magnitude difference
$\Delta M_{B}<1.5$.   An isolated galaxy furthermore cannot have any
neighbors within  $350h^{-1}$kpc$<r_{p}<700h^{-1}$kpc and $|\Delta
v|<1000$ km s$^{-1}$  with $\Delta M_{B}<0.75$; we relax our magnitude
cut at large $r_{p}$ because galaxies this  far apart will be less
dynamically associated.  Satellites are similarly  defined to be
galaxies with $r_{p}<350 h^{-1}$ kpc, $|\delta v|<500$ km s$^{-1}$ and
$\delta M_{B}>1.5$; i.e. satellites must be more than 1.5 magnitudes
fainter than the host galaxy they belong to.  These parameters define
sample 1 in Table 1 which  lists the search parameters used in this
analysis along with the number of  found satellites and several
derived host halo parameters.   We consider 7 different search
criteria and find that the results  presented are quite insensitive to
variations in these criteria; for  convenience we quote results from
sample 1 unless  otherwise noted.  Our choices of parameters ensure
that a satellite can be associated with one and only one host galaxy.
We put no restriction on morphological or spectral  type, but require
the isolated host galaxy to have $-22<M_{B}-5\log(h)<-21$ and
$0.7<z<1.0$.

\begin{figure}[t!]
\plotone{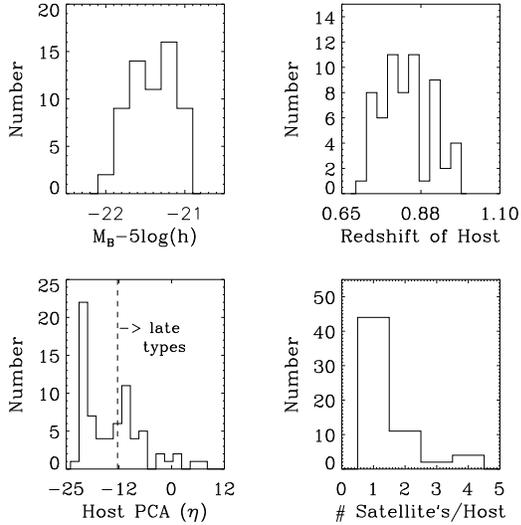}
\caption{Characteristics of isolated host galaxies.  
Upper left: absolute B-band magnitude of hosts.
Upper right: redshift distribtion.
Lower left: first PCA eigenvalue for host galaxy spectra.  The dashed line 
indicates the division between late types ($\eta>-13$) and early types 
($\eta<-13$).
Lower right: number of satellite galaxies found per host.}
\label{fig:hosts}
\end{figure}

\begin{figure}[t!]
\plotone{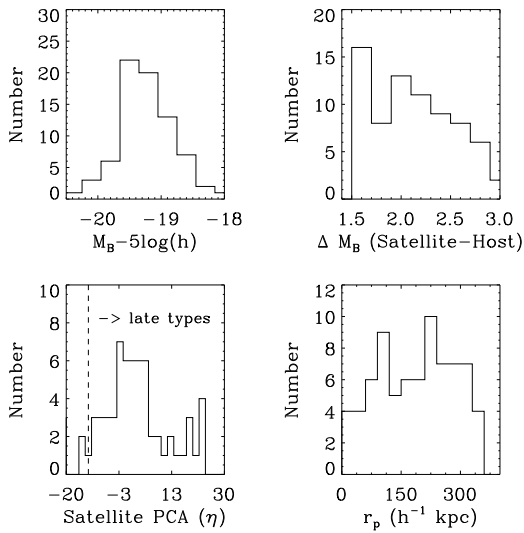}
\caption{Characteristics of satellite galaxy properties.  
Upper left: satellite galaxy absolute B-band magnitude.
Upper right: $\Delta M_{B}$ between host and satellite.
Lower left: satellite galaxy spectra type.  The dashed line indicates the 
division between late and early type spectra.
Lower right: projected separation (physical units) on the sky between 
satellite and host.  Slit collisions reduce the number of satellites found 
with $r_p<50 h^{-1}$ kpc.}
\label{fig:sats}
\end{figure}

For our chosen set of search parameters (Sample 1), we have identified
75 satellites around a total of 61 host galaxies at $0.7<z<1.0$ in the
DEEP2 data set.  Figures \ref{fig:hosts} and \ref{fig:sats} show
relevant characteristics of the satellite and host galaxies including
distributions in redshift, spectral-type, absolute magnitude,
satellite number per host, satellite distance from host, and $\Delta
M_{B}$ between the host and satellite.  We determine spectral types
using the principal component analysis of \citet{Madg03b} and use
their definition of $\eta=-13$ to separate early and late-type
galaxies.  Galaxy morphology and $(U-B)_0$ color correlate well with
this spectral classification of early and late types \citep{Madg03a}).
Satellites are found to have $\sim90\%$ late-type spectra, but due to
the DEEP2 survey selection effects mentioned above,  it is difficult
to determine if this is an intrinsic property of satellites around
bright isolated galaxies, or due to the $R$-band selection  of the
survey.  As we observe fainter galaxies (e.g.  satellites), early-type
galaxies become undetectable before late-type galaxies.

\begin{figure}
\plotone{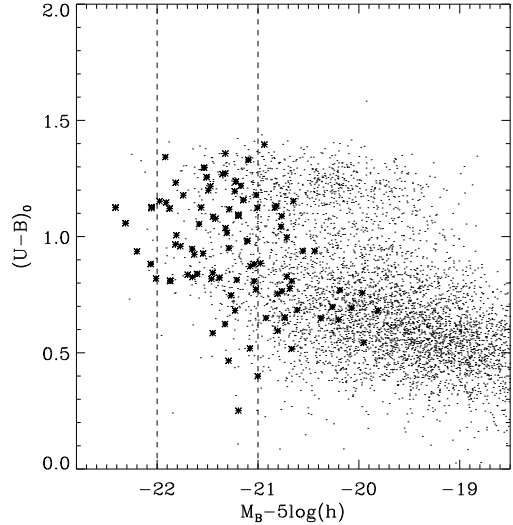}
\caption{Color-Magnitude diagram for all isolated host galaxies from 
sample 1 (stars) compared to a subsample of the first two seasons of DEEP2 
data (points).  The subsample was constructed to have the same redshift 
distribution as the host galaxies, thus reducing any potential redshift 
dependent selection effects.  Dashed lines indicate the magnitude range 
we have used in computing halo masses.  We classify galaxies with $(U-B)_0>$1 ``red'' and 
$(U-B)_0<$1 ``blue''.}
\label{fig:cmr}
\end{figure}

We find that approximately 60\% of host galaxies have early-type
spectra.  When we select a subsample of the entire available DEEP2
data set such that it has the same redshift and absolute magnitude
distributions as the host galaxies, we find that both sets of objects
have consistent fractions of early-type galaxies: 58\% for isolated
hosts and 63\% for the subsample.  This result is somewhat surprising;
one might have naively suspected that the majority of early-type
galaxies with $-22<M_B-5\log(h)<-21$ would reside in dense
environments and hence would not be identified as  isolated using our
search criteria.  To test the robustness of this result, we also use
$(U-B)_0$ color to test for any differences between isolated galaxies
and our reconstructed subsample.  Again we find that $\sim55\%$ of
isolated host galaxies are red ($(U-B)_0>$1), while $\sim50\%$ of the
subsample is red (see Fig. \ref{fig:cmr}).  There thus seems to be a
significant population of bright, red, early-type isolated galaxies
with satellites at $z\sim1$.  Our host galaxies have approximately
the same relative number of early and late spectral types as in P03's
sample; this  will allow for robust comparisons between halo masses
using SDSS data at $z\sim0$  and DEEP2 data at $z\sim1$.

\section{Halo Mass Estimation}\label{s:meth}

In this section we describe our method for obtaining dark matter halo masses
for isolated galaxies utilizing satellite galaxy kinematics.  Our approach is similar 
to previous work \citep[see e.g.][]{P03, brainerd03}, except for our maximum-likelihood
approach to deriving velocity dispersions which is more robust than previous methods when 
applied to small numbers of host-satellite pairs.

Schematically, we derive masses in the following way.  First we obtain a sample of 
isolated galaxies with associated satellites.  We then measure 
the line-of-sight velocity dispersion in several radial bins for a ``typical'' isolated
bright galaxy, using satellites as luminous tracers of the velocity
field.  In our sample, each isolated galaxy has at most three or four
satellites, but we can measure dispersions
by stacking the satellite-host pairs and thereby treating all satellites as
belonging to a single typical host galaxy.
In implementing this method we are assuming that similarly bright isolated galaxies reside
in similar halos.  We create an homogeneous host galaxy sample by
searching for satellites around galaxies with $-22<M_{B}-5\log(h)<-21$
and $0.7<z<1.0$.  As more data becomes available, it will be possible
to limit these criteria even further, yielding results for multiple
luminosity, redshift, color, and spectral type bins.

In order to determine line-of-sight velocity dispersions, we build distributions 
of the projected velocity difference between satellite and host ($\delta v$) 
in bins of projected radius from the host galaxy and fit for the 
dispersion in the distribution.  To convert the resulting 
velocity dispersion measurements into a mass, we make several assumptions, 
including the shape of the underlying dark matter potential.  We now 
describe this procedure in detail.

\subsection{Velocity Dispersion Measurement}

The difference between the host and satellite galaxy line-of-sight velocity,
$\delta v$, has a distribution that is well fit by
a Gaussian of zero mean.  In the absence of interlopers (see below), the satellite
velocity dispersion, $\sigma_{los}$ can simply be obtained from the width of a
Gaussian fit to the velocity distribution.   In order to
be able to detect variation in the velocity profile with radius, we
bin the satellites in projected radius, $r_p$, from the host galaxy.  We have chosen
bins such that the number of satellites per bin is roughly constant:
$30<r_{p}<180 $, $180<r_{p}<280$, and
$280<r_{p}<350$, in units of $h^{-1}$ kpc; these choices assure
similar errors from bin to bin.  

An important aspect of this analysis is the careful rejection of
``interlopers''; these are galaxies which meet the criteria for
satellite identification, but are in fact not dynamically associated
with the host.  Interlopers result from peculiar velocities which can,
in redshift space, scatter objects into our search cylinder.  Recent
local studies \citep[e.g. P03,][]{VDB04b} have found that about
20-30\% of putative satellites fall into this category.

Since interlopers are not 
physically associated with the host galaxy, we account for them 
by assuming that they will have a constant $\delta v$ distribution.  
Thus, we expect the observed $\delta v$ distribution be a combination of
flat (interloper) and Gaussian (true satellite) components.  We therefore 
fit a Gaussian plus constant distribution to the velocity
measurements within each $r_{p}$ bin.  If we ignore clustering
effects, which is reasonable since we are only probing isolated
systems, then one would expect the number density of interlopers to
simply scale with the search volume.  Although the
interloper fraction should be roughly constant in $\delta v$, that
constant should be different in bins of different radii.  This is 
another important motivation for measuring $\sigma_{los}$ in bins 
of $r_p$.

Unlike previous studies which fit Gaussian profiles
to velocity histograms, we determine the dispersion of the velocity
distribution using a maximum likelihood Gaussian-plus-constant fit to
the unbinned $\delta v$ data.  Specifically, our likelihood function:
\begin{equation}
L(a,\sigma_{los},i) = a + B e^{-\delta v_i^2/(2\sigma_{los}^2)},
\end{equation}
has two free parameters, a constant component (in satellites per km/s), $a$,
and the width of the Gaussian, $\sigma_{los}$.  The parameter $B$ is
chosen such that the integral of the probability density function of
the relative velocity distribution between host and
satellite (given $a$ and $\sigma_{los}$) over the allowed velocity 
range is one, and $\delta v_i$ is the $\delta v$ for the $i$th satellite-host
pair.  We maximize the summed logarithm of this likelihood function:
\begin{equation}
S(a,\sigma_{los}) = \sum_i ln(L)
\end{equation}
over a dense grid in $\sigma_{los}$ and $a$.

In Monte Carlo tests, this algorithm
agrees with method of fitting Gaussian distributions to velocity 
histograms for well-sampled data, but is much more robust in the
limit of small numbers of satellites.  This technique also provides 
an estimate of the error in the velocity dispersion measurement from 
the width of the likelihood peak.

\subsection{Halo Mass Determination}

In order to derive a host halo mass, we fit a theoretical velocity dispersion 
profile to the measured velocity dispersion points.  The theoretical profile is obtained 
via the following procedure.
We start by assuming an NFW \citep{NFW96, NFW97} density distribution

\begin{equation}
\frac{\rho(r)}{\rho^0_c} = \frac{\delta_c}{(r/r_s)(1+r/r_s)^2}
\end{equation}

($\rho \propto r^{-3}$ for large r) where $\rho^0_c$ is the present critical density, 
$r_s = r_{200}/c$, and 
\begin{equation}
\delta_c = \frac{200}{3} \frac{c^3}{ln(1+c) - c/(1+c)}
\end{equation}
where $r_{200}$, defined as the radius
where the mean interior density is 200 times the critical density.  
The concentration, $c$, can be viewed as a free parameter determining the 
shape of the NFW density profile.  In general the concentration is inversely related 
to the mass of a dark matter halo.

The Jeans equation is then used to relate the radial 
velocity dispersion, $\sigma_r$, to the gravitational potential, $\Phi$,

\begin{equation}
\frac{1}{\rho}\frac{d}{dr}(\rho\sigma_r^2)+2\beta\frac{\sigma_r^2}{r} = -\frac{d\Phi}{dr}
\end{equation}

and then we integrate along the line of sight

\begin{equation}
\sigma_{los}^2(r_p) = \frac{2}{\Sigma_M(r_p)}\int_{r_p}^\infty(1-\beta\frac{r_p^2}{r^2})\frac{\rho\sigma_r^2(r,\beta)r}{\sqrt{r^2-r_p^2}}dr
\end{equation}

where

\begin{equation}
\Sigma_M(r_p) = 2\int_{r_p}^\infty\frac{r\rho(r)}{\sqrt{r^2-r_p^2}}dr
\end{equation}

is the surface mass density (see \citet{LM01} for details of these
calculations).  In the above, $r$ is the radial distance and $r_p$ is, as usual, the 
distance projected on the sky.  The velocity anisotropy ($\beta \equiv
1-\sigma_{r}^2/\sigma_{\perp}^2$, where $\sigma_{r}$ is the radial
velocity dispersion and $\sigma_{\perp}$ is the velocity dispersion
perpendicular to the line of sight) must be assumed in the conversion
of the NFW density profile to a velocity dispersion profile; we use an
Osipkov-Merrit anisotropy, $\beta_{OM} = s^2/(s^2+s_a^2)$, with
$s_a=4/3$ and $s=r/r_{200}$.  \citet{VDB04b} and \citep{Mamon05} find 
that the line-of-sight velocity dispersion profile
depends only weakly on $\beta$, at the level of
a few percent, and hence we do not explore other parameterizations.  
We further need to assume the concentration; we use $c=10$,
 which is consistent with our fit to the mock catalog
velocity dispersion profiles (see $\S$~\ref{s:mocks}), and is in general in agreement 
with simulations of $\sim10^{12} M_{\sun}$ halos.  In $\S$~\ref{s:res} 
we show that our results concerning the evolution of the halo mass of isolated galaxies
are insensitive to the assumed concentration.  With these assumptions we
are left with only one free parameter, the normalization of the
velocity dispersion profile, which can be characterized by the
circular velocity at $r_{200}$, $V_{200}$.  We fit for $V_{200}$ via
$\chi^2$ minimization using the observed data points.  For the same
enclosed region, the interior mass ($M_{200}$) can be easily inferred
from $V_{200}$ for a given cosmology \citep[see][for
details]{NFW96,NFW97}.

\section{Results}\label{s:res}

\begin{figure}
\plotone{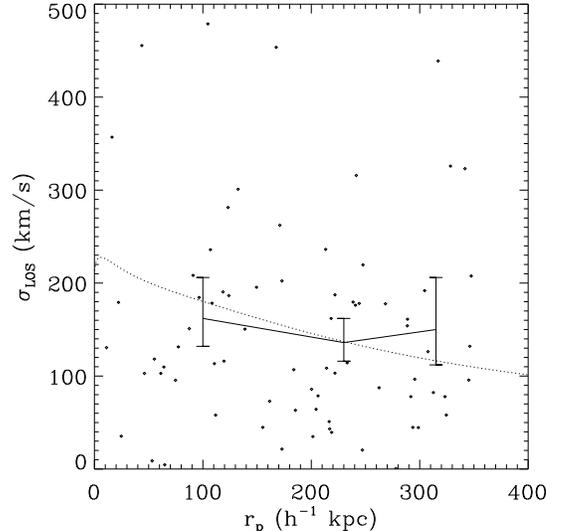}
\caption{Line-of-sight velocity dispersion profile for a typical
isolated  bright galaxy in the DEEP2 sample.  The line with error bars
is the profile  derived from the DEEP2 sample; the dotted curve is a
best fitting NFW   line-of-sight velocity dispersion profile ($c=10$
assumed).  The line with  errors is very stable over a wide range of
search parameters and variation  in the radial binning.  The
individual points are the $\Delta v$ and $r_{p}$  of the
satellite-host systems.  The best fit NFW profile (dashed line)
corresponds to a mass of $M_{200}=5.5 \times10^{12} h^{-1}M_{\Sun}$.}
\label{fig:vdp_deep}
\end{figure}

Using the maximum likelihood method outlined in $\S$~\ref{s:meth}, we measure a
velocity dispersion of $162^{+44}_{-30}$ km s$^{-1}$ for satellites
with $30<r_{p}<180 h^{-1}$kpc (median $r_{p}=110 h^{-1}$kpc),
$136^{+26}_{-20}$ km s$^{-1}$ for $180<r_{p}<280 h^{-1}$kpc (median
$r_{p}=230 h^{-1}$kpc), and $150^{+55}_{-38}$ km s$^{-1}$ for
$280<r_{p}<350 h^{-1}$kpc (median $r_{p}=320 h^{-1}$kpc) for isolated
galaxies with $-22<M_{B}-5\log(h)<-21$ (see Fig. \ref{fig:vdp_deep}).  
Errors on the velocity dispersion are derived from the maximum-likelihood fit.
These results are robust to changes in the search parameters and
radial binning; for the 7 different search criteria listed in Table 1,
our line-of-sight velocity dispersion measurements vary within
1$\sigma$ of the dispersions quoted above.  From Fig. \ref{fig:vdp_deep} it is clear that the
derived dispersion profile is consistent with nearly all popular halo
mass density profiles (e.g.  isothermal, NFW, Moore \citep{Moore98}),
though we use an NFW profile to derive masses.  What is important for this
analysis is the normalization of the velocity dispersion profile, not
the slope, as long as the slope at $V_{200}$ is shallow.

As outlined in $\S$~\ref{s:meth}, we fit velocity dispersion profiles derived 
from an NFW model to the measured velocity dispersion profile for DEEP2 
by minimizing $\chi^2$.  Our results imply a total mass, $M_{200}$, of 
$5.5^{+2.5}_{-2.0} \times10^{12} h^{-1}M_{\Sun}$ for a typical isolated 
galaxy with $-22<M_{B}-5\log(h)<-21$ and at least one satellite.  
When we vary the search criteria the measured mass varies by 
$\pm1\times10^{12} h^{-1}M_{\Sun}$, within our 1$\sigma$ errors (see Table 1).  As mentioned above, 
interlopers play a key role in this analysis.  From our best fitting Gaussian plus 
constant fit to the satellite $\delta v$ distribution, we find that the interloper fraction 
increases with increasing $r_p$, from $\sim7$\% at $r_{p}\sim110 h^{-1}$ kpc to $\sim30$\%
at $r_{p}=320 h^{-1}$ kpc.  These numbers are in good agreement with previous results from 
P03 and \citet{VDB04b}.

We can compare this derived mass to recent local measurements to measure evolution 
in the halo mass of isolated galaxies.  Unfortunately, we could
not implement the exact same search criteria as in P03 as that yields
only 30 satellites in the DEEP2 sample, too few to provide robust
results.  Thus, in order to make a fruitful comparison, 
we have taken the raw $\Delta v$ and $r_{p}$ measurements of
P03 (F. Prada 2004, private communication) and independently
determined the underlying halo mass using our own host galaxy
magnitude range and search criteria.  We note that our methods
accurately recover the masses inferred in P03's samples (see $\S$~\ref{s:intro})
when using their definitions and absolute magnitude intervals.

\begin{figure}
\plotone{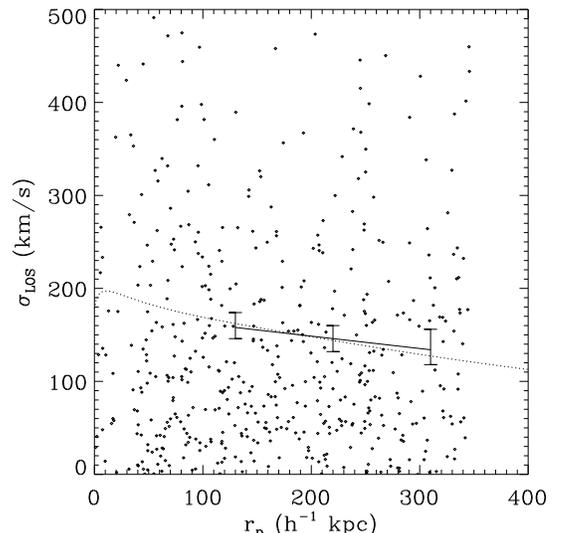}
\caption{As Fig. \ref{fig:vdp_deep}, but for SDSS satellite galaxies
from P03 (data provided by F. Prada).  The  best fitting NFW
line-of-sight velocity dispersion profile implies an average halo mass
$M_{200}$ of  $5.4^{+1.2}_{-1.0}\times10^{12} h^{-1}M_{\Sun}$.  The
lack of points past $r_p=350  h^{-1}$kpc is a result of the definition
of satellites in the P03 dataset.}
\label{fig:vdp_sdss}
\end{figure}

Taking $M^{*}_{B}(z\sim0)-5\log(h)=-19.45\pm 0.07$ \citep[][this value was 
converted from the $b_j$-band by assuming a median color of $B-V=0.21$]{Norberg02} and
$M^{*}_B(z\sim1)-5\log(h)=-20.6\pm 0.1$ \citep{willmer05}, we use 475 host
galaxies in the P03 sample with magnitudes $-21<M_{B}-5\log(h)<-20$
ensuring that the samples at $z\sim1$ and $z\sim0$ have similar host
galaxy magnitude ranges \emph{relative to $M^{*}_B$}.  We find
isolated galaxies at $z\sim0$ with $-21<M_{B}-5\log(h)<-20$ to have an
average halo mass of $5.4^{+1.2}_{-1.0}\times10^{12} h^{-1}M_{\Sun}$
when assuming an NFW mass density distribution with $c=10$ (see
Fig. \ref{fig:vdp_sdss}).

We can cast these results in terms of the dynamical mass to B-band 
light ratio, $M_{200}/L_B$.  Using the mean luminosity of host galaxies
at $z\sim 1$ of $M_{B}-5\log(h)=-21.5 \pm 0.1$ ($L_B=5.3\pm0.5\times 
10^{10} h^2 L_\Sun$) and at $z\sim 0$ of $M_{B}-5\log(h)=-20.5 \pm 0.1$ 
($L_B=2.1\pm0.1\times 10^{10} h^2 L_\Sun$), we find that the 
$B$-band mass-to-light ratio ($M_{200}/L_B$) is
increasing from $M_{200}/L_B=104\pm43$ $h M_\Sun / L_{\Sun,B}$ at $z\sim1$ to 
$M_{200}/L_B=257\pm54$ $h M_\Sun / L_{\Sun,B}$ at $z\sim0$, a factor of 2.5.  
Total (random and systematic) errors in our absolute magnitude measurements, 
including uncertainties in K corrections apart from the assumed cosmological 
parameters, are estimated to be below $0.1$ mag at the redshifts of interest 
\citep{willmer05}; hence errors in luminosities are negligible compared to the 
statistical uncertainties in the average $M_{200}/L_B$.

\section{Testing the Method}\label{s:mocks}

We use mock galaxy catalogs that have been constructed to match the
DEEP2 survey in order to test whether
we can accurately recover the halo mass of isolated galaxies using
satellites.  A complete description of the catalogs are given in
\citet{Yan04}; we give the relevant details here.  N-body simulations
of $512^3$ dark matter particles with a particle mass
$m_{part}=1.0\times10^{10}$ h$^{-1} M_{\Sun}$ were run in a
$\Lambda$CDM Universe using the TreePM code \citep{MWhite02}
in a periodic, cubical box of side length $256$ h$^{-1}$ Mpc.
Dark matter halos were identified by running a
``friends-of-friends'' halo finder with a minimum size of 8 dark matter
particles.  Galaxies down to $0.1 L^{*}$ are then assigned to 
individual dark matter particles via a halo model approach, in which 
both the number of galaxies within a halo and their luminosity function 
depends upon halo mass \citep{Yang03}.

We investigate the effects of
slitmask target selection \citep[see][for details]{Davis04} on the
derived velocity profile, thus testing our ability to recover the true
halo mass when these observational effects are included.  Target
selection may result in a galaxy being identified as isolated which is
not truly so, since we only have redshifts for approximately one-half
of the galaxies meeting the survey selection criteria.  Eventually,
photometric redshift estimates of galaxies without spectroscopy will
allow us to better constrain the number of truly isolated galaxies in
our sample.  Earlier work (P03) has investigated interloper
contamination using simulations of a single $10^{12} M_{\Sun}$ halo;
the large cosmological simulations we use also test the effects of
higher order clustering on the interloper problem.

\begin{figure}
\plotone{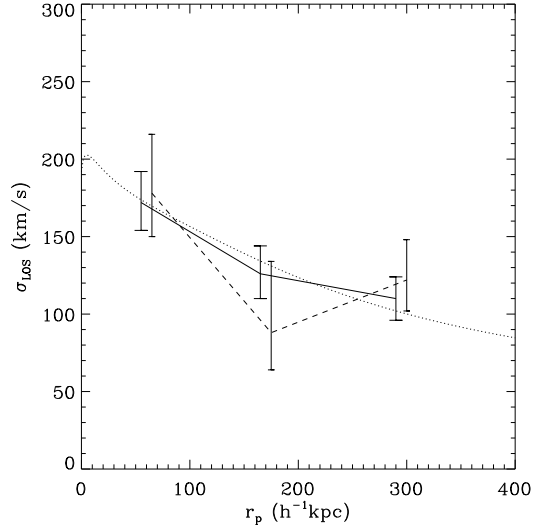}
\caption{The best fit NFW line-of-sight velocity dispersion profile (dotted line) to
the derived  profile for 480 satellites from mock catalogs (solid with 
error bars).  The mass associated with this NFW profile  is 
$3.8\times10^{12} h^{-1}M_{\Sun}$, which agrees quite well with the 
true average halo mass of these galaxies, $4.2\times10^{12} h^{-1}M_{\Sun}$. 
The dashed profile with error bars was derived from the mock catalogs after including
target selection effects.  The errors increase due to a smaller 
sample size and  contamination from galaxies falsely 
identified as isolated.  The best-fit NFW curve for this sample 
corresponds to a mass of $3.9\times10^{12} h^{-1}M_{\Sun}$, and again 
accurately recovers the true halo mass.  Note the different y-axis scales 
between this and Figs. \ref{fig:vdp_deep} and \ref{fig:vdp_sdss}.}
\label{fig:vdp_mocks}
\end{figure}

Fig. \ref{fig:vdp_mocks} compares a line-of-sight velocity dispersion profile derived
from an NFW density profile to profiles measured for isolated galaxies
in the mock catalogs with the same redshift and magnitude range and
search criteria as the data.  The mock catalogs employed here have
total volume that is comparable to the final DEEP2 sample.  The solid line is
the profile from the mock catalogs before we include the effects of
target selection.  The best-fit NFW profile (dotted line) is derived
as described in $\S$~\ref{s:meth}.  The halo mass associated with this profile
($M_{200}$) is $3.8^{+1.0}_{-0.87} \times10^{12} h^{-1}M_{\Sun}$.  We
can derive the true halo mass for these host galaxies by using the
simulations, since we know the number of dark matter particles in the
halo and hence can directly compute $M_{200}$.  The true halo mass
distribution for isolated galaxies in the mock catalogs conforms to a
log-normal distribution with 1$\sigma$ range $3.0\times10^{12}
h^{-1}M_{\Sun}$ to $1.0\times10^{13} h^{-1}M_{\Sun}$ with a peak of
$4.2\times10^{12} h^{-1}M_{\Sun}$.  The velocity profile fit therefore
recovers the true \emph{average} mass quite accurately, giving us
confidence that the algorithm used to fit the profile robustly
recovers the underlying halo mass.

The dashed line in Fig. \ref{fig:vdp_mocks} shows the velocity dispersion profile
derived from mock catalogs to which the DEEP2 target selection
algorithm (which will obtain spectroscopy of only $\sim 65$\% of eligible objects) 
and $70\%$ redshift completeness have been applied.
This profile is noisier both because of mistaken identification of
galaxies as isolated (due to their neighbors not being included in the
sample) and a 50\% decrease in the number of satellites due to the combined effects 
of target selection and redshift incompleteness.  
  
The mass associated with the best-fit NFW profile is
$3.9^{+1.8}_{-1.5} \times 10^{12} h^{-1}M_{\Sun}$.  The true masses of
the host halos conform to a log-normal distribution with 1$\sigma$
range $1.7\times10^{12} h^{-1}M_{\Sun}$ to $5.7\times10^{12}
h^{-1}M_{\Sun}$ and peak at $2.6 \times 10^{12} h^{-1}M_{\Sun}$.  When
we compare the isolated hosts found in the pre-target selection mock
catalogs to those found after target selection, we find that
$\sim30$\% of hosts in the mock catalogs after target selection are
not truly isolated, but only appear isolated because their companions
were not targeted for observation.  Fortunately, this level of
contamination does not seem to limit our ability to accurately recover
the underlying halo mass, as to first order it mimics the effects of
background interlopers and hence is accounted for by our interloper
correction.

\section{Discussion}\label{s:disc}

We find that isolated galaxies at $z\sim1$ have a similar halo
mass as isolated galaxies which are 1 magnitude fainter at $z\sim0$.
This implies that there has been little or no evolution in the halo
mass of isolated galaxies with magnitudes in the range $\sim
M^\ast_B$-0.5 to $\sim M^\ast_B$-1.5 even though $M_{B}^\ast$ has evolved by
$\sim1$ magnitude over this redshift range.  Our results are thus consistent with isolated
galaxies of a fixed luminosity relative to $M^\ast$ populating their dark
matter halos in a similar way from $z\sim1$ to $z\sim0$, a result
attained with an independent method by \citet{Yan03b}.  Phrased
differently, we find that the dynamical $B$-band mass-to-light ratio ($M_{200}/L_B$) is
increasing from $M_{200}/L_B=104\pm43$ $h M_\Sun / L_{\Sun,B}$ at $z\sim1$ to 
$M_{200}/L_B=257\pm52$ $h M_\Sun / L_{\Sun,B}$ at $z\sim0$, a factor of 2.5.  
This increase is attributable solely to the 1
magnitude decrease in the typical satellite host galaxy luminosity
from $z\sim1$ to $z\sim0$.  Assuming that the isolated galaxies found in DEEP2 at
$z\sim1$ passively evolve to the SDSS isolated galaxies at $z\sim0$,
our results imply that the ratio of baryonic mass to dark halo mass in
these galaxies has been constant for the last $\sim$8 billion years.

These results are insensitive to the various assumptions used
to calculate masses; when we instead assume an isothermal model 
for the dark matter density distribution, the values of the 
masses calculated for both the high and low redshift
samples change, but the consistency between these two values remains.
Specifically, for the isothermal model our isolated galaxy halo mass 
at $z\sim1$ within $r_{200}$ becomes $1.9\pm0.95\times10^{12}$ 
$h^{-1}M_{\Sun}$ while for isolated galaxies at $z\sim0$ 
the mass within $r_{200}$ changes to $2.0\pm0.56\times10^{12}$ 
$h^{-1}M_{\Sun}$ \citep[see][for the relevant relation between 
velocity dispersion and mass within $r_{200}$ in an isothermal model]{Bryan98}.
The same can be said for our other assumptions, including the
anisotropy required in the Jeans equation and the concentration for 
the NFW profile.  The value of the concentration parameter c is 
actually expected to be lower at high redshift than $z\sim0$ 
\citep{Bullock01}, but this, too, should not have a strong effect. 
For instance, if $c=5$ at $z\sim1$ instead of 10 \citep[following]
[who find that $c\propto(1+z)^{-1}$ for halos of the same mass]{Bullock01}, 
$M_{200}$ increases by $\sim1\times 10^{12}$ $h^{-1}M_{\Sun}$, well 
within the quoted errors.  It is essential for this comparison 
that masses at low and high redshift be calculated by the
same method and with the same assumptions, but the specific 
assumptions do not affect our overall conclusions.

Host-satellite kinematics in the local universe have also been studied
in the 2dF Galaxy Redshift Survey.  \citet{brainerd05} measure a 
velocity dispersion profile for $2L^\ast$ isolated galaxies and find that it 
falls from $\sim 200$ km s$^{-1}$ at $100 h^{-1}$ kpc to $\sim 160$ km s$^{-1}$ 
at $400 h^{-1}$ kpc, in good agreement with what we measure here for 
SDSS galaxies of similar luminosity.

Importantly, the halo mass derived from the data agrees with the mass 
derived from the mock catalogs for the same host magnitude range, 
$-22<M_{B}-5\log(h)<-21$.  Requiring such agreement can be used to 
set constraints on the way in which the number of galaxies in a dark 
matter halo and their luminosities depend on the underlying halo mass -- 
i.e., the halo occupation distribution and conditional luminosity function, which 
were used to create these mock catalogs.

Comparison to other methods for determining halo masses, such as galaxy-galaxy 
lensing, is complicated for a number of reasons.  First, to date galaxy-galaxy 
lensing studies have focused on $\sim L^\ast$ galaxies, while here we study 
a population of galaxies with mean luminosity $\sim 2.5L^\ast$.  Second, 
galaxy-galaxy lensing probes all of the mass along the line-of-sight, not 
just the mass around an isolated galaxy.  Third, we require satellites around 
isolated galaxies in order to determine the halo mass while galaxy-galaxy lensing 
can probe galaxies without satellites.  Halos with luminous satellites could 
potentially be systematically more massive than halos without luminous satellites.
Yet even with these potentially important differences, the evolutionary trend described 
in this paper is in good agreement with galaxy-galaxy lensing results which show that the halo 
mass of isolated $\sim L^\ast$ galaxies is approximately constant from $z\sim0.8$ to 
$z\sim0.15$ \citep{Wilson01, Hoekstra04, Kleinheinrich05}.
 
As mentioned in $\S$~\ref{s:res}, our results at $z\sim0$ are in good 
agreement with P03 when we use their definitions of host galaxies and masses.  
This good agreement is encouraging and implies that our mass reconstruction 
method is robust, as we do not exactly follow the mass estimation method 
outlined in P03.  Mass estimates at $z\sim0$ for isolated galaxies utilizing 
satellite kinematics seem to be converging.

It is important to keep in mind that, due to the small number of satellites 
found, this analysis has been applied to a host sample consisting of both 
early and late type galaxies, and is therefore mixing two different 
populations of galaxies.  We thus stress that these are initial results
requiring more data to untangle these sorts of complications.

Upon completion of the DEEP2 survey we will have a sample $\sim4\times$ 
larger than what was used for the present analysis, which will decrease 
our uncertainties on velocity dispersions by a factor of 2.  This 
will result in a similar decrease in errors on our halo mass estimate, 
allowing for much tighter constraints on both halo mass evolution and the 
halo model.  With the completed survey, we will be able to separate our host 
galaxies by spectral type, color, or redshift, allowing for more precise 
comparisons to local samples.

\acknowledgments
This project was supported in part by the NSF grants AST00-71198 and 
AST00-71048.  The DEIMOS spectrograph was funded by a grant from CARA 
(Keck Observatory), an NSF Facilities and Infrastructure grant (AST92-2540, 
the Center for Particle Astrophysics and by gifts from Sun Microsystems and 
the Quantum Corporation.  JN acknowledges support from NASA through Hubble 
Fellowship grant HST-HF-01165.01-A awarded by the Space Telescope Science 
Institute, which is operated by the Association of Universities for Research 
in Astronomy, Inc., for NASA, under contract NAS 5-26555.  Some of The data 
presented herein were obtained at the W.M. Keck Observatory, which is operated 
as a scientific partnership among the California Institute of Technology, 
the University of California and the National Aeronautics and Space 
Administration. The Observatory was made possible by the generous financial 
support of the W.M. Keck Foundation.  In addition, we wish to acknowledge 
the significant cultural role that the summit of Mauna Kea plays within 
the indigenous Hawaiian community; we are fortunate to have the opportunity 
to conduct observations from this mountain.

We would like to thank Francisco Prada for providing us with his 
satellite galaxy data.  C.C. would like to thank Chung-Pei Ma, Martin White, 
Francisco Prada, and Anatoly Klypin for enlightening conversations and for 
reading early drafts.


\begin{deluxetable}{rccccccccc}
\large
\tablecaption{search parameters and derived quantities for 7 samples.
$\delta$ refers to satellite selection criteria, while $\Delta$ refers 
to the parameters for host galaxy isolation.}

\tablehead{ \colhead{Sample} &  \colhead{$\delta M_{B}$ } &
\colhead{$\delta r_p$} &  \colhead{$\delta v$} &  \colhead{$\Delta
M_{B}$} &  \colhead{$\Delta r_p$} &  \colhead{$\Delta v$} &
\colhead{$N$ \tablenotemark{a}} &  \colhead{$V_{200}$
\tablenotemark{b}} &  \colhead{M \tablenotemark{c}}} \startdata 1 &
1.5 & 350 & 500 & 0.75 & 700 & 1000 & 75 & 410$^{+55}_{-60}$ &
5.5$^{+2.5}_{-2.0}$ \\ 2 & 1.0 & 500 & 750 & 0.75 & 1000 & 1000 & 160
& 410$^{+55}_{-60}$ & 5.4$^{+2.5}_{-2.0}$ \\ 3 & 1.5 & 500 & 750 & 1.0
& 1000 & 1000 & 52 & 405$^{+65}_{-70}$ & 5.3$^{+3.0}_{-2.3}$ \\ 4 &
1.5 & 350 & 700 & 1.0 & 700  & 1000 & 55 & 410$^{+75}_{-75}$ &
5.5$^{+3.6}_{-2.5}$ \\ 5 & 1.0 & 500 & 700 & 1.0 & 1000 & 1000 & 11 &
400$^{+70}_{-70}$ & 5.0$^{+3.1}_{-2.2}$ \\ 6 & 1.5 & 500 & 500 & 1.0 &
1000 & 1000 & 51 & 425$^{+65}_{-70}$ & 6.1$^{+3.3}_{-2.6}$ \\ 7 & 1.5
& 350 & 500 & 1.0 & 500  & 1000 & 82 & 425$^{+60}_{-60}$ &
6.1$^{+2.9}_{-2.2}$ \\ \enddata \tablenotetext{a}{Total number of
satellites with host magnitude  $-22<M_B-5\log(h)<-21$.}
\tablenotetext{b}{$V_{200}$ measured in km s$^{-1}$}
\tablenotetext{c}{$M_{200} / 10^{12} h^{-1}M_{\Sun}$}
\end{deluxetable}

\end{document}